\documentclass[aps,prl,twocolumn,showpacs,groupedaddress,preprintnumbers,amsmath,amssymb]{revtex4}

\usepackage {amsmath,amssymb,epsfig,color}
\usepackage {graphicx}% Include figure files
\usepackage {dcolumn}% Align table columns on decimal point
\usepackage {bm}% bold math

\begin{document}

\title{LATTICE PARAMETERS MODIFICATION OF PARA-DICHLORBENZOL NANODIMENSION FILMS}

\author {M.A.~Korshunov}
 \email {mkor@iph.krasn.ru}
 \affiliation {L.V. Kirensky Institute of Physics, Siberian Branch of Russian Academy of Sciences, 660036 Krasnoyarsk, Russia}

\date{\today}

\begin{abstract}
Influence of decrease of the sizes of crystallites of 
para-dichlorbenzol on Raman spectrums is studied. Lattice oscillations 
frequencies are calculated. On this basis the modification of lattice 
parameters for particles smaller than 5$\mu $ and an incremented 
modification of nanodimension sized particles are studied. Magnification of 
lines intensities are related to transmitting oscillations and orientation 
oscillations frequency decrease is found.
\end{abstract}

\maketitle

As it is scored in a number of operations of a lattice of [1] parameters 
nanodimension particles constructed on atoms differ from parameters of a 
lattice of crystals of the greater size. Thus for one crystal the sizes of a 
lattice can be incremented, for others will decrease. The modification of 
physical properties because of resizing crystallites for the structures 
constructed of organic molecules, apparently, should be observed at their 
greater size, than for the structures constructed of atoms. Organic crystals 
find the increasing application in the molecular electronics engineering (at 
making crystals, and in units of storage). Therefore as will lead parameters 
of a lattice for organic molecular crystals at decrease of particle sizes up 
to nanodimensions, represents practical interest. For learning this problem 
it has been carried out examination of particles from 5$\mu $ and below 
$\sim $1$\mu $ organic matter by a method of a Raman effect of light of 
small frequencies. This method has been selected, as spectrums of the 
lattice oscillations are rather sensitive to modify structure of a crystal. 
As the object of examination the molecular crystal of para-dichlorbenzol 
which is the good modelling object has been selected. Para-dichlorbenzol was 
studied by various methods. It crystallizes in space group P2$_{1}$/a$_{}$with two molecules in cell [2].

Samples were prepared by a following method. On a glass plate (an 
integumentary glass) the studied crystal has been raised out of dust. A film 
derivated consist, apparently, from separate crystallites. After a 
sputtering the film was evaporated up to the necessary width. It has been 
prepared three exemplars by width from 30$\mu $, $\sim $5$\mu $ and less 
1$\mu $. The film consists of separate crystallites. From above the received 
film of the necessary width has been covered by other integumentary glass, 
for decrease of transpiration. After that Raman spectrums of small 
frequencies have been received.

\begin{figure}
\includegraphics[width=0.7\linewidth]{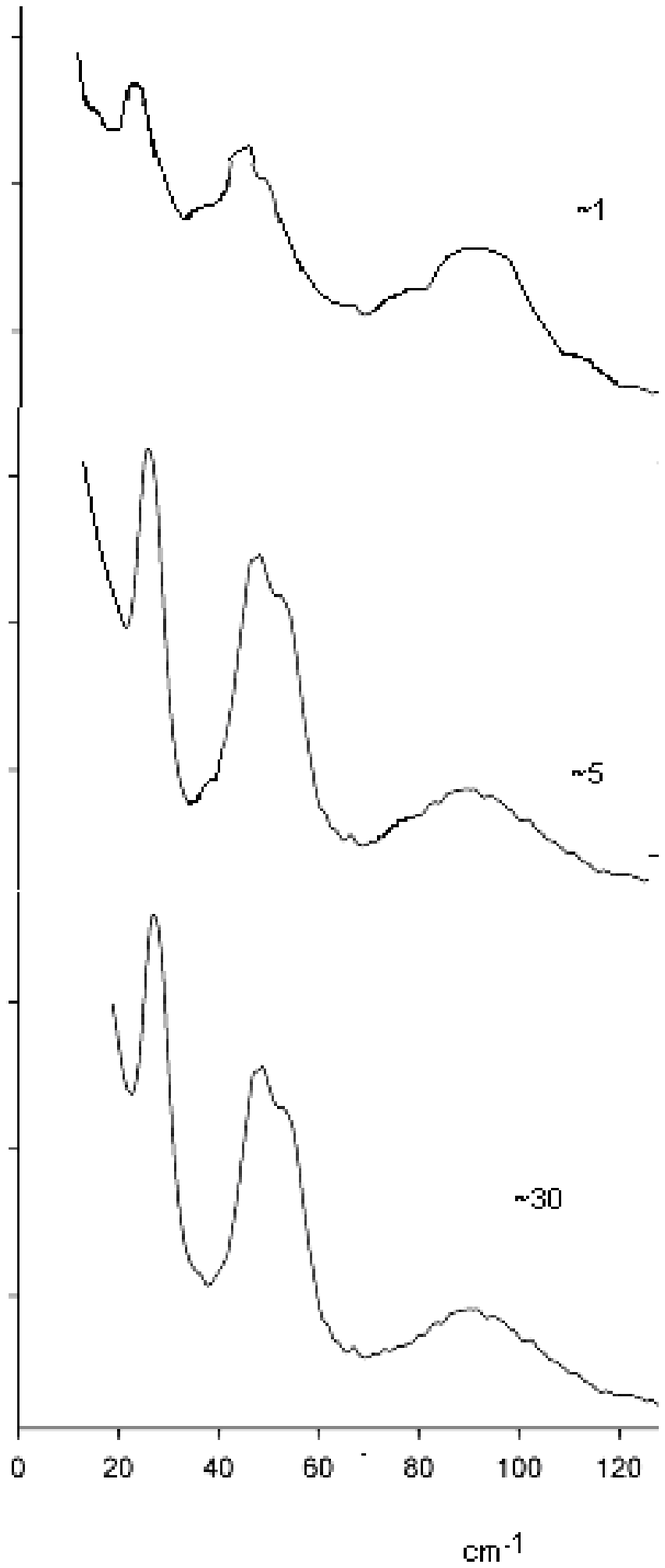}
\caption{A spectrum of the lattice oscillations of para-dichlorbenzol at 
resizing crystallites $\sim $30$\mu $, $\sim $5$\mu $ and less 1$\mu $.}
\label{fig1}
\end {figure}

In Fig.~\ref{fig1} the spectrums of the lattice oscillations obtained for films of 
studied width of para-dichlorbenzol are shown. In a spectrum of the lattice 
oscillations of a single crystal it should be observed six intensive lines 
of the molecules caused by rotational effects around of moments of inertia. 
As we see, there is a decrease of the value of frequencies of similar 
spectral lines. In a spectrum detrusion of frequencies depending on the 
sizes of crystallites is observed, the it is less particle size, the 
detrusion is more.

\begin{table*}
\caption{Modification of orientation oscillations frequencies with a 
modification of a film width.}
\begin{tabular}
{|p{68pt}|p{68pt}|p{68pt}|p{68pt}|p{68pt}|p{68pt}|p{68pt}|}
\hline
$\sim 1\mu ,$cm$^{ - 1}$& 
22.5& 
43.5& 
45.5& 
49.5& 
91.0& 
100.0 \\
\hline
$\sim 5\mu ,$cm$^{ - 1}$& 
25.8& 
45.5& 
47.5& 
53.5& 
92.0& 
100.5 \\
\hline
$\sim 30\mu ,$cm$^{ - 1}$& 
27.0& 
47.0& 
48.0& 
55.0& 
92.5& 
101.0 \\
\hline
\end{tabular}
\label{tab1}
\end{table*}

Intensity of lines of the molecules caused by rotational hunt effects around 
of an axis with the greatest moment of inertia also decreases with decrease 
of particle sizes. The same lines have also the greatest detrusion of 
frequencies. As we see, the modification of frequencies of lines even if it 
is a film and its size is observed varies in one direction. This film it is 
possible to consider as quasi a two-dimensional crystal. In a spectrum with 
decrease of width of a film appearance of additional lines of small 
intensity is observed. Intensity of these lines increases.

For an explanation of observable modifications of frequencies of lines 
calculations of spectrums of the lattice oscillations by a Dyne method [3] 
have been done. And calculations on transformation of parameters of a 
lattice at resizing crystallites. Minimization on energy was done for 
various layouts of molecules and their amount.

At matching the calculated spectrum with experimental it is discovered, that 
parameters of a lattice with decrease of the sizes of crystallites for 
para-dichlorbenzol are incremented. If to guess, that decrease of value of 
frequencies is linked only to decrease of the sizes of crystallites, how 
show calculations should be observed more significant decrease of 
frequencies for rotational oscillations linked with the least moment of 
inertia. The part of lines of small intensity can be caused by transmitting 
oscillations (these are lines in the field of 36 ñm$^{ - 1}$ and 70 ñm$^{ - 
1})$ which intensity at decrease of width of a film is incremented. 
Appearance of additional lines can be caused by the several reasons that 
demands the further examination.

Thus, it is shown that at examination of films of organic crystals coming 
nearer to nanosized particle by a method of a Raman effect of light decrease 
of value of frequencies of the lattice oscillations is observed. As show 
calculations it is caused by magnification of parameters of a lattice of 
para-dichlorbenzol. As magnification of intensity of lines linked with 
transmitting oscillations is observed. The modification of physical 
properties because of resizing crystallites for the structures constructed 
of other organic molecules, probably, has other character.

\end{document}